\documentclass[a4paper,11pt]{article}
\usepackage{pos}
\usepackage{natbib}

\title{Supernovae Ia and Gamma-Ray Bursts together shed new lights on the Hubble constant tension and cosmology}
\ShortTitle{SNe cosmology and GRBs}

\author[1,2,3,4,5]{M. G. Dainotti}
\author[6,7]{B. De Simone}
\author[8,9]{G. Montani}
\author[10]{E. Rinaldi}
\author[11,12]{M. Bogdan}
\author[13]{K. M. Islam}
\author[14]{A. Gangopadhyay}

\affiliation[1]{Division of Science, National Astronomical Observatory of Japan, 2-21-1 Osawa, Mitaka, Tokyo 181-8588, Japan}
\affiliation[2]{The Graduate University for Advanced Studies (SOKENDAI), Shonankokusaimura, Hayama, Miura District, Kanagawa 240-0115}
\affiliation[3]{Space Science Institute, 4765 Walnut St Ste B, Boulder, CO 80301, USA}
\affiliation[4]{Nevada Center for Astrophysics, University of Nevada, 4505 Maryland Parkway, Las Vegas, NV 89154, USA}
\affiliation[5]{Bay Environmental Institute, P.O. Box 25 Moffett Field, CA, California}
\affiliation[6]{Department of Physics, University of Salerno, Via Giovanni Paolo II, 132 I-84084 Fisciano (SA), Italy}
\affiliation[7]{INFN, Sezione di Napoli, Gruppo collegato di Salerno, Italy}
\affiliation[8]{ENEA, Fusion and Nuclear Safety Department, C.R. Frascati, Via E. Fermi 45, Frascati, I-00044 Rome, Italy}
\affiliation[9]{Physics Department, “Sapienza" University of Rome, P.le Aldo Moro 5, I-00185 Rome, Italy}
\affiliation[10]{Interdisciplinary Theoretical \& Mathematical Science Program, RIKEN (iTHEMS), 2-1 Hirosawa, Wako, Saitama 351-0198, Japan}
\affiliation[11]{Department of Mathematics, University of Wroclaw, plac Uniwersytecki 1, 50-137 Wrocław, Poland}
\affiliation[12]{Department of Statistics, Lund University, Box 117, SE-221 00 Lund, Sweden}
\affiliation[13]{Department of Physics, University of Constantine 1 - RN79, Constantine, Algeria}
\affiliation[14]{Department of Physics, Faculty of Science and Engineering, Konan University, 8-9-1 Okamoto, Kobe, Hyogo 658-8501, Japan}

\emailAdd{m.dainotti@nao.ac.jp}
\emailAdd{bdesimone@unisa.it}
\emailAdd{giovanni.montani@enea.it}
\emailAdd{erinaldi.work@gmail.com}
\emailAdd{malgorzata.bogdan@uwr.edu.pl}
\emailAdd{islam.islam.khadir6@gmail.com}
\emailAdd{anjasha@hiroshima-u.ac.jp}

\abstract{\textbf{Abstract}\\
The $\Lambda$CDM model is the most commonly accepted framework in modern cosmology. However, the local measurements of the Hubble constant, $H_0$, via the Supernovae Type Ia (SNe Ia) calibrated on Cepheids provide a value which is in significant disagreement, from 4 to 6 $\sigma$, with the value of $H_0$ inferred from the Cosmic Microwave Background (CMB) observed by \textit{Planck}. This disagreement is the so-called \textit{Hubble constant tension}. To find out the reason for this discrepancy, we analyze the behaviour of the $H_0$ in the Pantheon sample of SNe Ia through a binning approach: we divide the Pantheon into 3 and 4 bins ordered with redshift ($z$), and for each of them, we estimate the $H_0$. After the $H_0$ estimation, we fit the $H_0$ values with a decreasing function of $z$, finding out that $H_0$ undergoes a slow decreasing trend compatible with the evolution scenario in 2.0 $\sigma$. Such a behaviour could be explained by hidden astrophysical biases or the evolution with $z$ for the SNe Ia parameters. If not, the $f(R)$ modified gravity theories could be invoked to alleviate or solve the $H_0$ tension. Together with SNe Ia, more astrophysical probes such as quasars (QSO) \citep{Colgain,DainottiQSO} and Gamma-Ray Bursts (GRBs), which are much more distant than SNe Ia, are needed to tackle the $H_0$ tension. In the realm of GRB-cosmology, one of the most promising correlations is the fundamental plane relation, which is among the luminosity at the end of the plateau emission, its rest-frame duration, and the peak prompt luminosity \citep{Dainotti2016,Dainotti2017,Dainotti2020}. In the context of applying this relation as a cosmological tool, we also compute how many GRBs must be gathered to reach the same precision as the SNe Ia. Since we are about two decades away from reaching such precision, we also attempt to find additional correlations for the GRBs associated with SNe Ib/c  that could be exploited to standardize the class of GRB-SNe Ib/c in the future. We find a hint of a correlation between the GRBs' end-of-plateau optical luminosity and the SNe's rest-frame peak time, suggesting that the GRBs with the most luminous optical plateau emission are associated with SNe with the most delayed peaks in their light curves. So far, it is the fundamental plane relation to be the most promising candle for exploring the high-$z$ universe.}

\ConferenceLogo{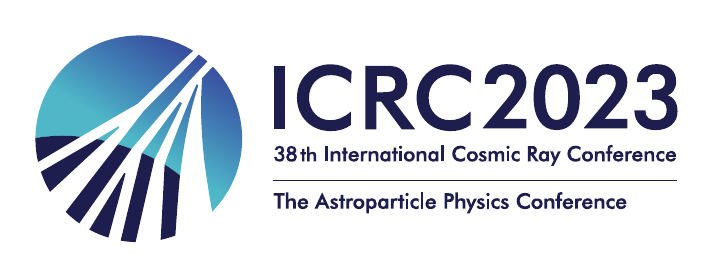}

\FullConference{%
38th International Cosmic Ray Conference (ICRC2023)\\
  26 July - 3 August, 2023\\
  Nagoya, Japan}

\begin{document}
\maketitle

\section{Introduction}
\vspace{-0.45cm}
Modern cosmology relies on the widely accepted $\Lambda$CDM, or, standard cosmological model.
Despite its success, the $\Lambda$CDM model faces several unresolved issues, notably the Hubble constant tension, or $H_0$ tension. This tension manifests as a significant discrepancy, in the range of 4 to 6 $\sigma$, between the value of $H_0$ derived from observations using SNe Ia calibrated on Cepheids and the one obtained from Cosmic Microwave Background (CMB) radiation.
Addressing this problem requires using reliable tools for cosmological analysis, which should actually be \textit{standard candles}, namely astronomical objects whose luminosity is known or can be derived through established relations. Together with SNe Ia, which are among the best standard candles so far discovered, the GRBs play a crucial role since they are observed up to $z=9.4$ (a redshift much greater than SNe Ia) and they can extend the Hubble diagram further. The GRBs with plateau emission, in particular, have proven to be standardizable for cosmological applications. The plateau is a flattening of the GRB LC observed after the prompt emission in multiwavelengths (from $\gamma$-rays to optical and, occasionally, in radio). This component is not only of astrophysical interest, given that it can be explained through the fallback of materials in a black hole \cite{Rowlinson2014} or the spinning down of a magnetar \cite{Rea2015}, but also allows the existence of the correlation between the X-ray luminosity at the end of the plateau emission, $L_X$, and its rest frame time, $T^*_X$ \citep{Dainotti2008,Dainotti2013,Dainotti2020,Dainotti2022ApJS,Dainotti2022MNRAS,Dainotti2022PASJ}, and the correlation between $L_X$ and the 1s peak prompt luminosity, $L_{peak}$ \citep{Dainotti2011,Dainotti2015}. The combination of these two correlations reveals the fundamental plane relation among $L_X$, $T^*_X$, and $L_{peak}$ \citep{Dainotti2016,Dainotti2017,Dainotti2020}. 
This correlation has been applied as a reliable cosmological tool \citep{Cao2022a,Cao2022b,Dainotti2022PASJ,Dainotti2022MNRAS,Dainotti2022MNRAStool,Bargiacchi2023} and can read distances much larger than the ones observed by SNe Ia observed up to $z=2.26$ \citep{Rodney2016}.
However, GRBs simulations show that we can achieve the same precision on the cosmological parameter $\Omega_M$ as the current SNe Ia samples in less than twenty years from now \citep{Dainotti2022MNRAS}. Thus, finding new correlations in the GRB-SNe could significantly help their future standardization. In Section \ref{sec:h0tension}, we report the analysis of the $H_0$ tension through the Pantheon sample binning with redshift \citep{Dainotti2021SNe,Dainotti2022SNe}, while in Section \ref{sec:grbsne} we present the search for new correlations in the GRB-SNe events \citep{Dainotti2022GRBSNe}. We provide our conclusions in Section \ref{sec:conclusions}.
\vspace{-0.4cm}
\section{Part 1. The Pantheon sample and the $H_0$ tension}\label{sec:h0tension}
\begin{figure}[h]
    \centering
    \includegraphics[scale=0.33]{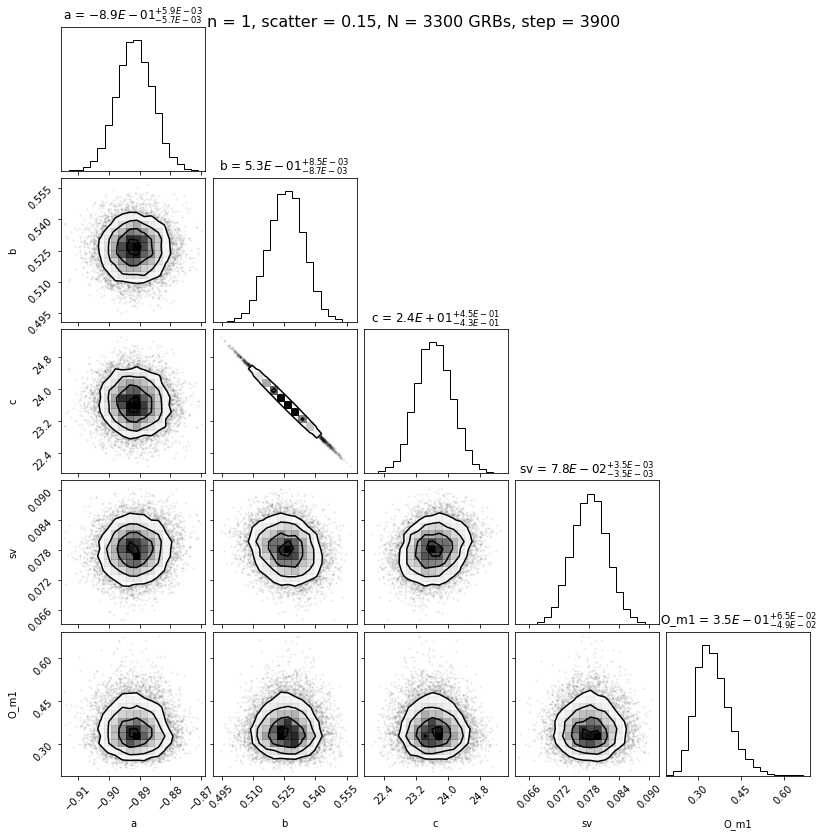}
    \vspace{-2ex}
    \caption{\textbf{Left.} The estimation of $\Omega_{M}$ through the simulation of 3300 GRBs that mimic the platinum sample properties. The $a,b,c,sv$ are the parameters of the fundamental plane while $O\_m1$ indicates the $\Omega_{M}$.}
    \label{fig:simulations}
\end{figure}

\vspace{-0.4cm}
A binning approach is employed to investigate the $H_0$ tension within the Pantheon sample \citep{Scolnic2018}, which consists of 1048 SNe Ia. The sample is split into 3 and 4 bins equally populated bins of SNe Ia ordered by $z$. A Monte Carlo Markov Chain (MCMC) analysis is conducted in each of these bins, using a $\chi^2$ test minimization comparing the observed and the theoretical SNe distance moduli. We vary $H_0$ in the $\Lambda$CDM and the $w_{0}w_{a}$CDM models \citep{CPL} in each bin, then after we obtain the values of $H_0$ in each bin, we then fit the $H_0$ values with the following function: $H_0(z)=\mathcal{H}_0/(1+z)^\epsilon$, where $\mathcal{H}_0=H_0(z=0)$ and $\epsilon$ is the Hubble constant evolution coefficient, where $H_0=73.5\,km\,s^{-1}\,Mpc^{-1}$. The results in \citep{Dainotti2021SNe} show that $H_0$ undergoes a slow decreasing evolution with $z$: the parameter $\epsilon$ is $\sim 10^{-2}$ and is compatible with zero in 2.0 $\sigma$. In the $\Lambda$CDM case, the values of $\epsilon$ are the following: $0.009\pm0.004$ in the 3 bins division (it is not compatible with zero up to 2 $\sigma$, namely, $\epsilon / \sigma_{\epsilon}= 2.0$).
In \citep{Dainotti2022SNe}, we enlarge the parameters space and consider two free parameters at the same time: $H_0,\Omega_{M}$ (where $\Omega_{M}$ is the total matter density of the universe) in the $\Lambda$CDM model and $H_0,w_{a}$ in the $w_{0}w_{a}$CDM model (considering that the equation of state is $w(z)=w_{0} + w_{a}*z/(1+z)$ and $w_{0}=-1.009$). Then, we divide the Pantheon sample into only three bins to prevent the statistical fluctuations from dominating the results. Differently from \citep{Dainotti2021SNe}, where the priors for $H_0$ are uniform and have a wide parameter space ($60\,km\,s^{-1}\,Mpc^{-1}<H_0<80\,km\,s^{-1}\,Mpc^{-1}$), we here adopt the choice of Gaussian priors for $H_0,\Omega_{M},$ and $w_a$ considering the values we expect from the normal distributions of each parameter: the central values used are $\Omega_{M}=0.298\pm0.022,H_0=70.393\pm1.079,w_{a}=-0.129\pm0.026$ and we draw priors from the distributions extended up to 2 $\sigma$. 
Furthermore, we add the contribution of Baryon Acoustic Oscillations (BAOs, \citep{BAOs}) in the SNe Ia bins. We confirm the slowly decreasing behaviour of $H_0(z)$, with $\epsilon$ still $\sim 10^{-2}$ and being compatible with zero only in 5.8 $\sigma$, a much more significant level than before. The highlighted behaviour of $H_0(z)$, if not caused by unseen astrophysical biases or $z$-evolution effects of SNe Ia parameters (e.g. the drift with $z$ for the SNe Ia stretch parameter discussed in \citep{Nicolas2021}), could find a natural explanation in the paradigm of modified $f(R)$-gravity in the so-called Jordan frame \citep{Sotiriu,SchiavoneMontaniBombacigno2023}.
In fact, in such a formulation, the standard gravitational field is non-minimally coupled to a scalar field, whose potential term summarizes the effect of non-Einsteinian gravity. The idea underlying the interpretation of the data consists in the dynamical rescaling of the Einstein constant by the scalar field dynamics. In particular, we can model the $z$-dependence of the scalar field so that the phenomenological behaviour of $H_0(z)$ is reproduced. The two basic equations for the isotropic Universe dynamics provide the specific form of the potential term as a function of $z$ and, hence, are also expressed via the scalar field itself. The $f(R)$-gravity that emerges from this analysis corresponds to a viable model in the sense that it is tachyon-free and fulfills the observational constraint at $z\simeq 0$. Thus, we can conclude that a possible interpretation of the Hubble tension, viewed here in terms of the $H_0(z)$ trend, could be naturally accounted for in the framework of a revised dynamical formulation of the underlying geometrodynamics, well-reconciliated at low $z$ with the standard $\Lambda$CDM model. 
After exploring the possible theoretical explanation for this $H_0$ trend, we still need to add additional probes at higher redshift to confirm this trend. Thus, we further investigate the use of the GRB fundamental plane and we estimate the number of GRBs needed by using the fundamental plane to reach SNe Ia's precision in computing the $\Omega_{M}$ cosmological parameter. Mimicking the variables of the fundamental plane relation for GRBs in X-rays \citep{Dainotti2016,Dainotti2017,Dainotti2020}, we simulate 3300 GRBs that follow the properties of the platinum sample \citep{DainottiPlatinum}. Leaving free to vary the parameters of the GRBs fundamental plane ($a,b,c,sv$) and $\Omega_{M}$ we obtain, for 3300 GRBs $\Omega_{M}=0.350\pm0.057$, see Figure \ref{fig:simulations}. This number shows that we have reached precision twice as high on $\Omega_{M}$ than the one of \citep{Conley2010} ($\Omega_{M}=0.18 \pm 0.10$) and we are close to the precision of \citep{Betoule2014} which is $\Omega_{M}=0.295\pm0.034$. Since the corresponding optical 2D and 3D relation \citep{Dainotti2020,Dainotti2022ApJS} are more constraining for cosmological applications according to \citep{Dainotti2022MNRAS} simulations, if we use the optical sample, the light curve reconstruction \citep{Dainotti2023LCR} to halve the fitting parameters errors of the fundamental plane, and the machine learning to double the sample of GRBs with $z$, then we can reach the precision of \citep{Betoule2014} in just a decade from now in 2032 \citep{Dainotti2022MNRAS}.

\begin{figure}[h]
    \centering
    \includegraphics[scale=0.12]{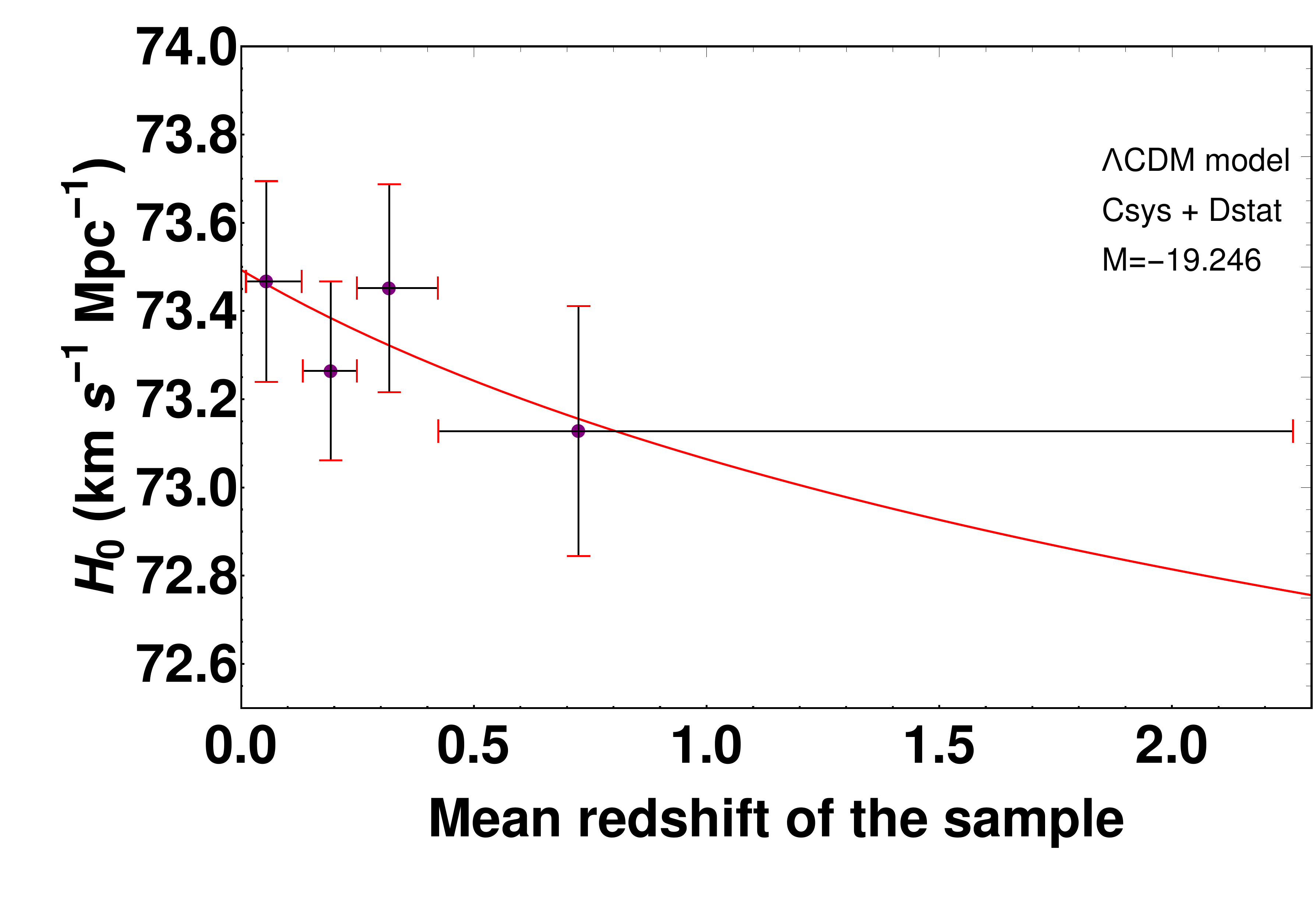}
    \includegraphics[scale=0.12]{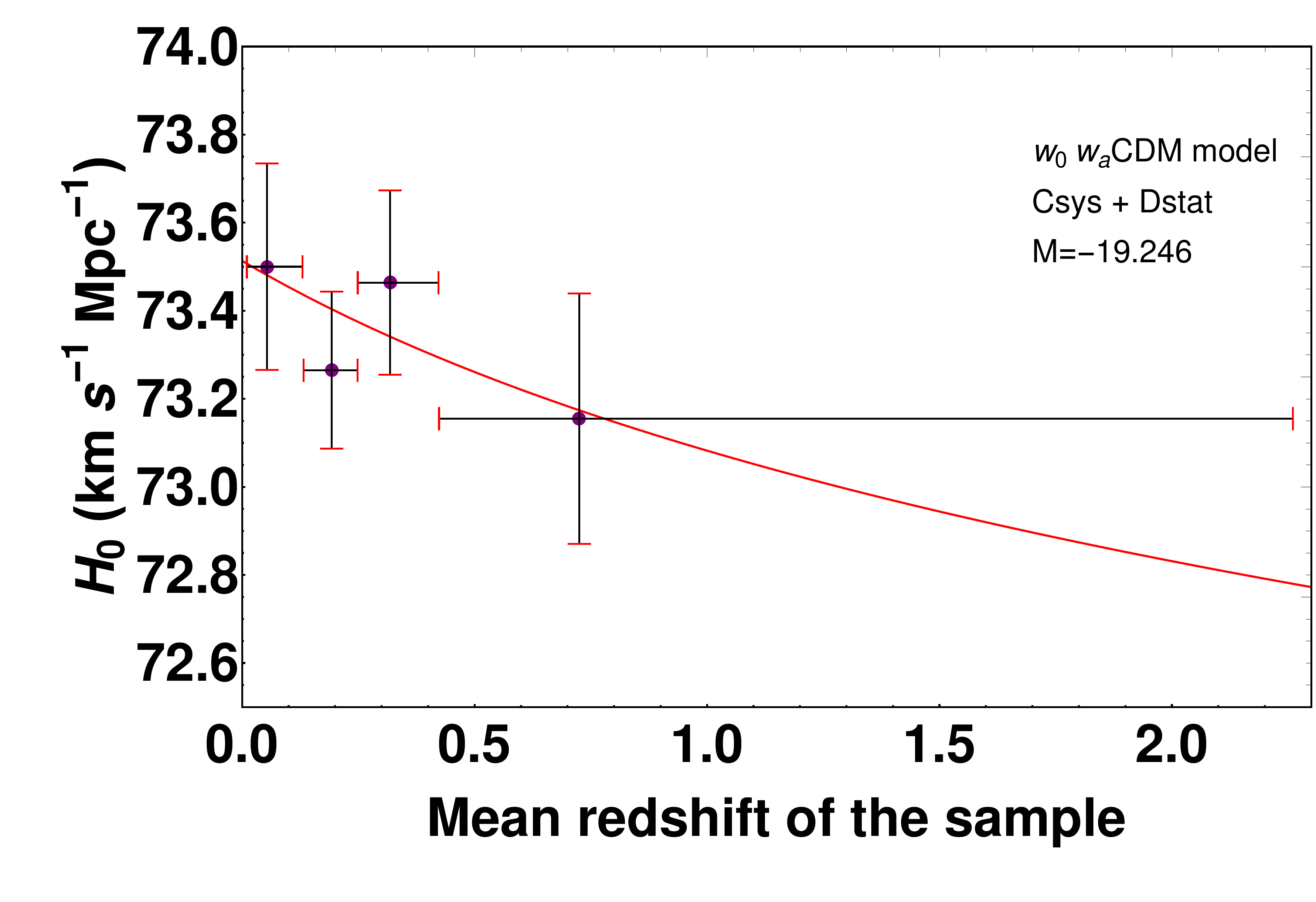}
    \vspace{-3ex}
    \caption{The $H_0(z)$ plot in the 4 bins case \citep{Dainotti2021SNe}.  \textbf{Left.} The $\Lambda$CDM model case is shown. \textbf{Right.} The $w_{0}w_{a}$CDM model is present.}
    \label{fig:4bins}
\end{figure}

\vspace{-0.45cm}
\section{Part 2. Exploring the correlations for the GRB-SNe events.}\label{sec:grbsne}
\begin{figure}
    \centering
    \includegraphics[scale=0.248]{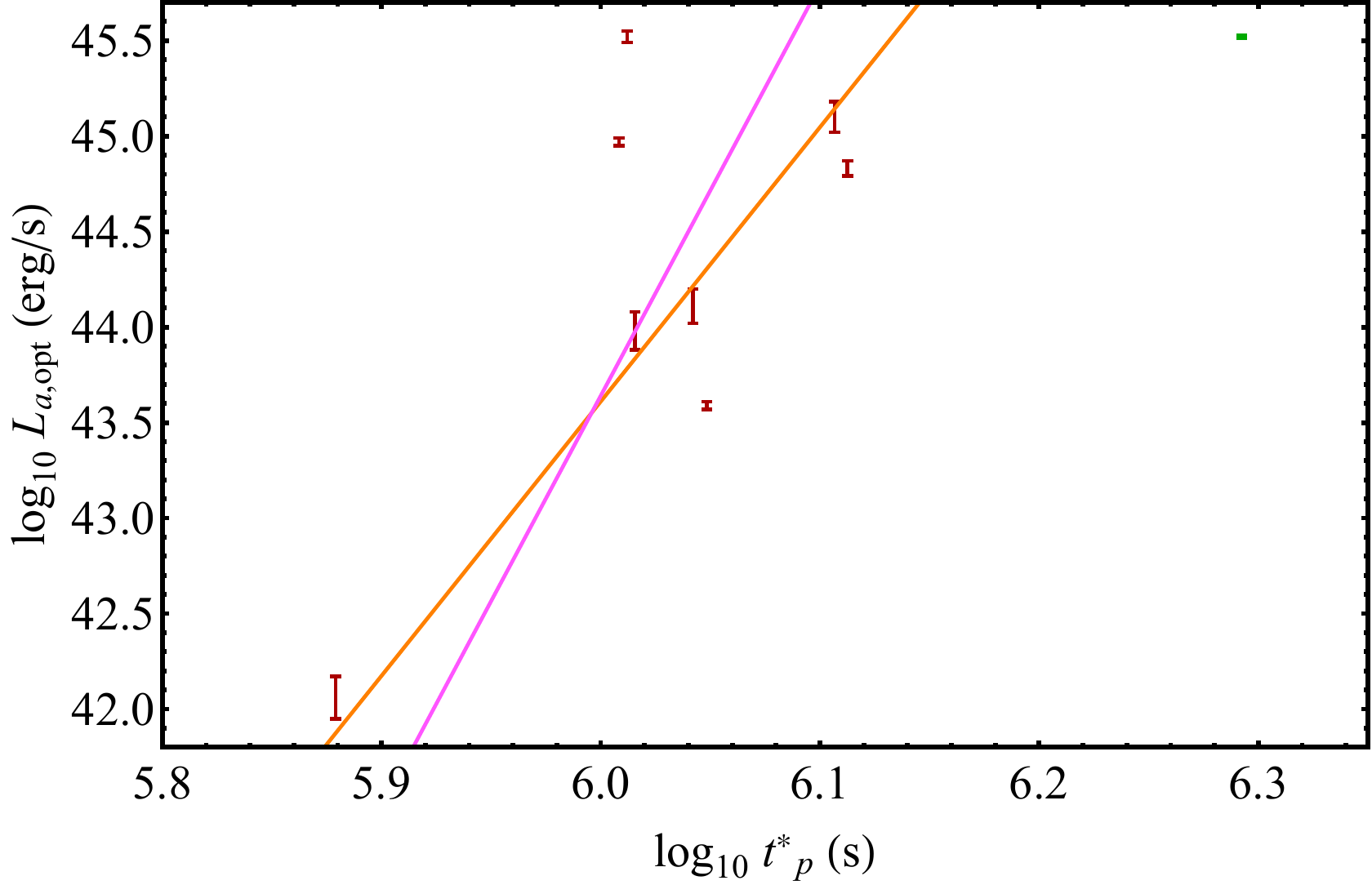}
    \includegraphics[scale=0.22]{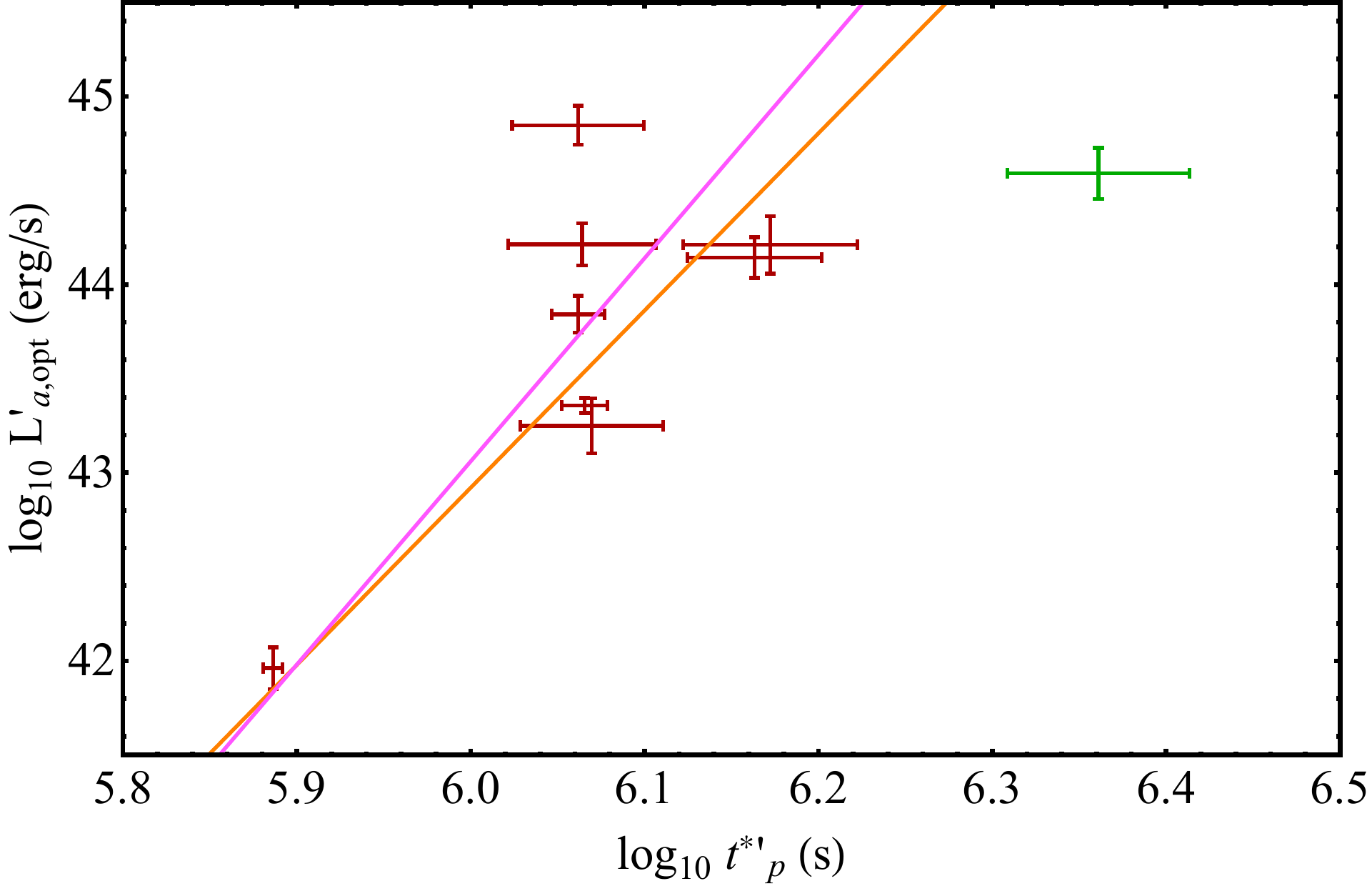}
    \caption{$L_{a,opt}-t*_p$ relation plot and fitting \citep{Dainotti2022GRBSNe}. \textbf{Left.} Fitting with the variables uncorrected through the EP method. \textbf{Right.} Variables are corrected with the EP method. The red points correspond to A,B graded GRB-SNe, while the green point refers to an E-graded GRB-SN event.}
    \label{fig:grbsnefit}
\end{figure}

\vspace{-0.35cm}
To tackle the $H_0$ tension, it becomes crucial to search for new correlations also at low-redshift so that they can be anchored to the SNe Ia. We here discuss the correlations in the realm of the GRB-SNe and, in future, standardize this class of events. GRB-SNe are a class of Long GRBs (LGRBs, with a typical duration greater than 2 seconds) associated with SNe Ib/c and explained with the collapse of massive stars. They are observed up to $z \sim 1$ and are classified with A, B, C, D, and E grades \citep{Hjorth2012}: A, B) = strong spectroscopic hints about the presence of an associated SN; C) the SN bump in the GRB LC late-times, is typical of the GRB-SNe observations; D) a significant bump, but not typical of the GRB-SNe class; E) a non-significant bump is observed in the LC. Here, we present the work performed in \citep{Dainotti2022GRBSNe} where, using a complete sample of GRB-SNe from April 1991 to February 2021 and introducing the GRB-SN afterglow properties for the first time in the correlations research, they investigate the presence of linear 2-dimensional relations among GRB and SNe parameters. From the initial 106 possible GRB-SN connections, 35  associations based only on spatial coincidence, and two shock breakout events are excluded (GRBs 060218A, 080109A). The latter two events are explained in the \textit{failed jet} scenario and differ from the other GRB-SNe. We then apply the Efron \& Petrosian (EP) method \citep{EP1992} to remove the redshift evolutionary effects from the GRB and SNe parameters. The symbol ($'$) denotes the variables corrected for evolution. After correcting for evolution, we test all the possible 91 bidimensional correlations among GRB and SNe parameters, and we apply the following two metrics: (1) Pearson coefficient $|r_P| \geq 0.5$ and its $p$-value $P_P \leq 0.05$; (2) Spearman coefficient $|r_S| \geq 0.5$ and its $p$-value $P_S \leq 0.05$. As a result, we find a probable correlation between the GRBs at the end of the plateau optical luminosity ($L'_{a,opt}$) and the SNe rest-frame peak time ($t'*_p$). The weighted fitting relation is: $L'_{a,opt}=(9.43\pm1.97)t*'_p + (-13.60\pm11.89)$, with $r_P=0.71,P_P=0.03$ (see Figure \ref{fig:grbsnefit}). This implies that the most the SNe are delayed the most luminous the GRB plateau emissions is. 

\vspace{-0.4cm}
\section{Conclusions and future perspectives}\label{sec:conclusions}
\vspace{-0.4cm}
\noindent A decrease of $H_0$, present in the Pantheon sample, could be either attributed to unknown astrophysical biases or to the $z$-evolution of SNe Ia parameters or to the fact that we need to invoke $f(R)$ theories of gravity. The study of the $H_0$ through the new Pantheon+ \citep{PantheonPlus} will cast more light on this trend and its nature, if still present. We have shown attempts to uncovering new correlations within the GRB-SNe associations. A probable relation between GRBs' plateau luminosity in the optical and the SNe's rest-frame peak time is highlighted.
Nevertheless, this correlation still needs future confirmation since the GRB-SNe are relatively rare events (around 3 GRB-SNe observed each year), and we need more of these transients to investigate further their nature. Another perspective comes from the investigation of multidimensional correlations within the GRB-SNe class. In conclusion, being the scatter on the fitting parameters significant for the $L'_{a,opt}-t*'_p$ with the current data, the fundamental plane Dainotti relations confirm to be one of the best candidates for the use of GRBs as standardizable candles \citep{Dainotti2016,Dainotti2017,Dainotti2020}. In fact, the simulations of GRBs drawn from the fundamental plane relation in the optical show that we can expect to reach the same precision of SNe Ia in the estimation of the $\Omega_M$ cosmological parameter by 2032 through machine-learning and light curve reconstruction techniques \citep{Dainotti2023LCR}.

%
%
%

\end{document}